\documentclass[authoryear]{emulateapj}

\newcommand{\hr}{\mbox{$^h$}}
\newcommand{\mt}{\mbox{$^m$}}
\newcommand{\st}{\mbox{$^s$}}
\newcommand{\degree}{\mbox{$^{\circ}$}}
\newcommand{\am}{\mbox{\arcmin}}
\newcommand{\as}{\mbox{$^{\prime\prime}$}}

\newcommand{\kms}{\mbox{km\,s$^{-1}$}}

\newcommand{\um}{$\mu$m}
\newcommand{\cm}{cm$^{-1}$}

\def\lsim {$\rlap{\raise.4ex\hbox{$<$}}\lower.55ex\hbox{$\sim$}\,$}

\newcommand\mir{mid-infrared}
\newcommand\nir{near-infrared}
\newcommand{\lsun}{\mbox{L$_\odot$}}
\newcommand{\msun}{\mbox{M$_\odot$}}
\newcommand{\av}{\mbox{$\rm A_V$}}
\newcommand{\vlsr}{\mbox{$\rm v_{LSR}$}}
 
\newcommand{\hh}{H$_2$}

\newcommand{\hho}{H$_2$O}
\newcommand{\nhhh}{NH$_3$}

\newcommand{\neii}{[Ne\,II]} 
\newcommand{\ariii}{[Ar\,III]}
\newcommand{\siv}{[S\,IV]} 
\newcommand{\arv}{[Ar\,V]} 

\shorttitle{W51 IRS\,2 IONIZED JET}
\shortauthors{LACY ET AL.}

\begin{document}

\title{W51 IRS\,2: A MASSIVE JET EMERGING FROM A MOLECULAR CLOUD INTO AN H\,II
REGION}

\author{John H. Lacy\altaffilmark{1},
Daniel T. Jaffe\altaffilmark{1},
Qingfeng Zhu\altaffilmark{2},
Matthew J. Richter\altaffilmark{3},
Martin A. Bitner\altaffilmark{1},
Thomas K. Greathouse\altaffilmark{4},
Kevin Volk\altaffilmark{5},
T. R. Geballe\altaffilmark{5},
and David M. Mehringer\altaffilmark{6}}

\altaffiltext{1}{Department of Astronomy, The University of Texas at Austin,
Austin, TX 78712; lacy, dtj, mbitner@astro.as.utexas.edu}
\altaffiltext{2}{Center for Imaging Science,
Rochester Institute of Technology, Rochester, NY 14623; qxzpci@cis.rit.edu}
\altaffiltext{3}{Department of Physics, University of California,
Davis, CA 95616; richter@physics.ucdavis.edu} 
\altaffiltext{4}{Lunar and Planetary Institute, Houston, TX 77058,
greathouse@lpi.usra.edu}
\altaffiltext{5}{Gemini Observatory, Hilo, HI 96720; kvolk,
tgeballe@gemini.edu}
\altaffiltext{6}{davemehringer@yahoo.com}

\begin{abstract}

We have mapped \neii\ (12.8\,\um) and \siv\ (10.5\,\um) emission
from W51 IRS\,2 with TEXES on Gemini North,
and we compare these data to VLA free-free observations and
VLT \nir\ images.
With 0.5\as\ spatial and 4\,\kms\ spectral resolution we are able to
separate the ionized gas into several components:
an extended H\,II region on the front surface of the molecular cloud,
several embedded compact H\,II regions, and a streamer of high
velocity gas.
We interpret the high velocity streamer as a precessing
or fan-like jet, which has emerged from the molecular cloud
into an OB star cluster where it is being ionized.

\end{abstract}


\section{INTRODUCTION}

Most star formation, across the IMF, takes place in massive
star-forming cores \citep{lada03}.  
The disruption of these cores as high
mass stars form terminates the formation of low and intermediate
mass stars and is a major contributor to the dynamical energy
in the dense ISM.  While much of this disruption results from
the ionization of HII regions and their subsequent expansion,
it is becoming increasingly evident that dynamically
significant amounts of supersonic gas are present during
the very early phases of core dissipation \citep{depree04}.
In this paper we present observations of a jet containing a significant
fraction of the kinetic energy of an expanding H\,II region.

W51 IRS\,2, which includes the compact H\,II region W51d and the
maser complex W51 NORTH, lies within the H\,II region complex G49.5-0.4
in W51A, a very luminous
region of massive young stars and molecular and ionized gas
\citep{bieging75,genzel82,carpenter98}.
The distance to W51 NORTH has been determined from \hho\ maser
proper motion studies
to be 6-8 kpc \citep{genzel81,schneps81,imai02},
which is consistent with the far kinematic distance for the
molecular cloud velocity of 57\,\kms.
We will use a distance of 7 kpc, at which 1 pc subtends 30\as .

Radio free-free and recombination line observations of W51 have been made by
\citet{martin72}, \citet{gaume93}, and \citet{mehringer94}.
The 3.6\,cm free-free image of IRS\,2 shown in \citet[Figure 19]{mehringer94}
is from observations using all four VLA configurations, which preserves
both large and small scale structure.
We show a larger portion of that image in Figure 1 (Plate x)
and the IRS\,2 region in Figure 2a.
Various names have been given to the emission peaks in IRS\,2.
We will refer to the free-free peaks labeled a-f in Figure 2a as IRS\,2a-2f,
and the compact infrared continuum source as 2i.
Our IRS\,2a and 2b have been called W51d$_1$, our 2d is W51d, and our
2e is W51d$_2$.

\citet{okamoto01} observed infrared fine-structure line emission
of \neii\ (12.8\,\um), \ariii\ (9.0\,\um), and \siv\ (10.5\,\um) from
the ionized gas, and concluded that the gas is ionized by radiation
from stars of spectral type O8-O9.
At 7 kpc, the ionized gas in IRS\,2 requires a Lyman continuum luminosity
of $6.5 \times 10^{49}$ sec$^{-1}$ \citep{martin72}, and
the luminosity of IRS\,2 is $2-4 \times 10^6$ \lsun\ 
\citep{erickson80,jaffe85}.

Near-infrared observations show a cluster of young stars coincident with
IRS\,2 containing 2-8 times as many ionizing stars as the trapezium
cluster \citep{goldader94,hodapp02}.
Stellar colors indicate an extinction of \av $\sim$25 mag.
An H-band adaptive optics image with 0.05\as\ resolution,
with the 3.6\,cm contours overlaid, is shown in Figure 1b.
This image was taken with NAOS-Conica on the VLT Yepun as part of
program 71.C-0344(A), and was obtained from the VLT archive.
Most of the point sources are stars in the IRS\,2 cluster.
The diffuse emission correlates rather well with the 3.6\,cm
emission, although it appears to be cut off on the northern side,
with the northeastern half of the IRS\,2d missing,
probably because of extinction.
The diffuse H-band emission probably is dominated by
H\,I Brackett series lines.

The W51 NORTH maser group includes \hho, OH, SiO, and \nhhh\ masers.
The majority of the \hho\ maser peaks are in the `dominant center' group
\citep[`m' in Fig. 2a]{schneps81,imai02},
which is centered on a more compact
cluster of SiO masers \citep{eisner02} and on the peak of the thermal
\nhhh\ emission \citep{ho83}.

\section{OBSERVATIONS}

We observed W51 IRS\,2 during the Texas Echelon Cross Echelle Spectrograph
(TEXES)
Demonstration Science run on the 8-m Gemini North telescope in July 2006.
TEXES is a high resolution cross-dispersed \mir\ (5-25\,\um) spectrograph
capable of 0.4\as\ and 3\,\kms\ resolution on Gemini \citep{lacy02}.
All data from the Science Demonstration run are available from the
Gemini archive at http://archive.gemini.edu.
Members of the astronomical community are welcome to propose to use TEXES.
We expect it to be available in alternating semesters on Gemini
and the IRTF.

The observations presented here were made
with a 0.6\as\ slit width and 0.15\as\ sampling along the slit.
W51 IRS\,2 was mapped at three spectral settings:
10.5\,\um, including the 951.43\,\cm\ \siv\ line;
12.8\,\um, including the 780.42\,\cm\ \neii\ line;
and 13.1\,\um, including the 763.23\,\cm\ \arv\ line.
The maps were made by stepping the 4\as -long N-S slit in 0.15\as\ steps
across 12\as\ in R.A.
Overlapping scans were made to cover 9\as\ in DEC at
10.5\,\um\ and 7\as\ at 12.8\,\um.
An additional ten points were observed at the beginning of each scan
to measure the sky background.

A map of the 12.8\,\um\ continuum, from an 0.6\,\cm\ interval
next to the \neii\ line, is shown in Figure 2b.
The \siv\ and \neii\ line intensities, integrated over Doppler ranges
of \vlsr\ = -80 to +80\,\kms , with continuum subtracted,
are shown in Figures 2c and 2d.
The \neii\ image is similar to the radio free-free image (Fig. 2a),
with peaks seen at the positions of IRS\,2a-c and possibly 2e-f.
A bright peak is seen on the south edge of IRS\,2d, but the northeast
side of 2d is weak or absent in all of our maps.
The \siv\ map is somewhat different, with IRS\,2a and 2b
absent and 2d even more cut off.
With these bright sources suppressed, 2c and the extended emission
covering most of the region mapped are more prominent in \siv.
The \arv\ line was undetected, and must be fainter than \neii\ by
a factor of at least 30.

Our continuum maps are consistent with those of \citet{okamoto01},
who derive a silicate feature optical depth of $\tau_{9.7} \approx 2$
toward most of the ionized gas,
implying a reddening between our lines of $\rm E_{10.5-12.8} \approx 1$.
Given the rather uniform reddening,
except toward IRS 2i, where it is much larger,
the differences between the \neii\ and \siv\ maps must be due primarily
to excitation or density differences.
Of these two lines, \neii\ normally dominates in gas excited by stars
of spectral type O8 and later, and \siv\ dominates in gas excited by
earlier type stars.
However, \siv\ has a critical density for thermalization more than an
order of magnitude less than \neii , decreasing the \siv / \neii\ ratio
in gas with $n_e > 10^4$ cm$^{-3}$.
IRS\,2a and 2b, and the south edge of 2d, with [S\,IV]/[Ne\,II] $< 1/3$,
are of relatively low excitation (or high density),
whereas 2c and the extended emission have [S\,IV]/[Ne\,II] $\approx$ 3.

The fits format data cubes are available from the Gemini archive.
We show several position-velocity diagrams from the \siv\ cube
in Figure 3 and broad channel maps in Figure 4.
Movies of the P-V diagrams and maps are available on-line.
The locations of the P-V cut lines are shown in Figure 2c.
The Doppler shift of the positive velocity emission is close to that
of the IRS\,2 molecular cloud.
There is a trend that the positive velocity emission toward
center of the map is more redshifted than that near the edges.
The emission near the center is redshifted from the ambient molecular
gas by a few \kms.
That near the edges is blueshifted by $\sim$20\,\kms.
The most striking aspect of the P-V diagrams is the wide range in Doppler
shift seen near and south of IRS\,2c.
The 2c emission ridge is blueshifted from the ambient gas
by $\sim$110\,\kms.
This emission was seen by \citet{mehringer94} in the H92$\alpha$ line,
but was not resolved spatially, and the bridge between
the blueshifted and ambient emission was not apparent.
This `bridge' emission makes it clear that the blueshifted gas is not just
superimposed along the line of sight, but is interacting with the
ambient gas.
This is especially clear in P-V cut c, in which the Doppler
shift of the ambient gas shifts to the blue where it connects into
the bridge gas, and there is a gap in the ambient
gas in the region where the bridge is strongest.

\section{INTERPRETATION}

Gas at Doppler shifts near that of the molecular cloud is seen
throughout the region observed.
Given the presence of a cluster of massive young stars centered on this
region, we identify the exteded gas as the surface of the
cloud which is ionized by stellar UV.
The moderate extinction to the gas and stars indicates that they
lie near the front side of the cloud.
The pattern of Doppler shifts ranging from
just redward of the ambient cloud velocity
near the center of the region and blueward toward the edges is
consistent with a flow along a bowl-shaped front surface of the cloud.
\citet{zhu05} show that the motion of the ionized gas in several compact
H\,II regions can be explained in this way, and they suggest that such a
flow can result, at least in part,
from the pressure gradient in ionized gas compressed by
the ram pressure of stellar winds.
The [S\,IV]/[Ne\,II] ratio indicates that the gas is ionized by stars of
spectral type O8 or earlier.

The large reddening of IRS\,2i suggests that it
lies within the molecular cloud behind the ionized gas.
In addition, 2a and 2b must be separate
from the rest of the ionized gas, given their relatively low excitation,
indicating ionization by $\sim$O9 stars.
However, this gas
must not be very deeply embedded and could even be in a foreground cloud.
Several of the brightest stars in Figure 1b lie near ionized
gas peaks, suggesting that the star cluster is still partially embedded.

The explanation for the highly blueshifted gas in IRS\,2c is not so clear,
but the presence of `bridge' gas at intermediate velocities
connecting the ambient velocity and blueshifted emission in Figure 3
suggests that the blueshifted gas is in some sort of a jet
that passed through the ionized surface of the molecular cloud.
The jet must be neutral (to be unseen in the ionic lines)
while inside of the cloud and probably is ionized by
stellar UV when it emerges from the cloud.
The presence of high velocity \hho\ and SiO maser peaks 1\as\ 
south of 2c (near the Figure 3b cut line) suggests that the
jet is a part of the same outflow as that in which the masers form.
If this is the case, the jet must have precessed by $\sim$90\degree\ since
the gas we see was emitted, since the masers are moving near the plane of
the sky in an approximately E-W bipolar pattern, whereas the ionized jet
is to the north of the maser center and has a substantial velocity
component toward us.
However, the ionized jet
may not be the same as the one responsible for the `dominant center'
maser peaks.
There are several other groups of masers in IRS\,2,
which suggests that there are several outflow centers.
We may be able to localize better the origin of the outflow we see from
the information in the \siv\ data cube.
Just south of 2c there is a gap in the cloud surface gas
shown in Figure 4a,
which could be caused by the passage of the jet through
the surface H\,II region.
If it is, the source of the jet must lie farther to the south,
and so to the south of the maser center.
Observations of a neutral jet tracer, such as shocked \hh ,
could localize the source and test this interpretation.

The E-W elongation of the blueshifted emission could be a result of the
precession of a narrowly collimated jet, or the gas could have been
emitted in a fan.
In either case, the relatively small variation in the Doppler shift
at the peak of the emission along the blueshifted jet suggests that
the angle of the jet to our line of sight does not vary by a large
amount along its length.
This indicates that the distance along our line of sight to the source
of the jet is at least several times greater than the 0.13 pc
length of the blueshifted emission.
We can estimate the time since the jet was emitted from the ratio of 
the distance to the ionized jet from its source to the observed velocity.
This gives a time of $>$1200 yr.

The extent of the intermediate velocity `bridge' gas in Figures 3 and 4
is something of a puzzle.
The narrowness of the blueshifted ridge indicates that the jet responsible
for it is well collimated, at least in declination.
But the extent of the bridge emission is almost as great in declination
as it is in right ascension.
Two explanations seem plausible.
As the surface of the jet is photoionized, it would be expected to expand,
and it might slow down by interaction with the surrounding medium.
Alternatively, the jet might carry along with it material from the
H\,II layer on the surface of the molecular cloud.
\citet{williams04} has suggested that a turbulent magnetized medium
should act as a visco-elastic fluid.
It would be interesting to ask whether this effect could be strong
enough to account for the acceleration of the bridge gas.

We can estimate the mass of the ionized gas in the blueshifted emission
from the ionic line fluxes.
\citet{okumura01} derived an electron density in the ionized gas
in IRS\,2 of $10^5 - 10^6$ cm$^{-3}$ from the ratio of near-infrared
[Fe\,III] lines.
At that density, the \siv\ line
should be thermalized, making its flux proportional to mass.
We then calculate the mass of ionized gas in the jet to be
$3 \times 10^{-3}$\,\msun\ assuming all of the S is S$^{+3}$,
and S/H is solar.
Since some of the gas may be in other ionization states
(in fact the core of the jet most likely is neutral, at least when
it emerges from the molecular cloud), this mass estimate is a lower limit.
Assuming the ionized part of the jet includes gas emitted over 500 yr,
we obtain a bipolar mass flow of $> 1.2 \times 10^{-5}$ \msun\,yr$^{-1}$
and an energy flow of $>$10\,\lsun.
For comparison, molecular outflows from massive young stars range upward
from $3 \times 10^{-5}$ \msun\,yr$^{-1}$ \citep{garay99}.

The high velocity streamer in W51 is similar to the irradiated Herbig-Haro
jets seen by \citet{reipurth98} and \citet{cernicharo98},
but has an emission measure a factor of $10^5$
greater than those of the irradiated jets in the Orion nebula
\citep{bally01} and a mass flow rate a factor of at least $10^4$
greater than those jets.
It more closely resembles the HH 80-81 bipolar jet \citep{marti93},
which comes from a massive star.
The HH 80-81 jet has much greater extent than that in W51, but is not
as luminous in ionic emission, likely due to the lack of as strong of
a source of ionizing radiation.
It is highly collimated and shows some evidence of precession, although
not as much as we suggest for the W51 jet.

\acknowledgements

We thank the Gemini staff, especially John White, for their outstanding
help in making TEXES work on Gemini North.
We also thank Jim Umbarger for assistance with the figures.
The development of TEXES was supported by grants from the NSF and the
NASA/USRA SOFIA project.
The modification of TEXES for use on Gemini was supported by Gemini
Observatory.
Observations with TEXES were supported by NSF grant AST-0607312.
This work is based on observations obtained at the Gemini Observatory,
which is operated by the Association of Universities for Research in
Astronomy, Inc., under a cooperative agreement with the NSF on behalf of the
Gemini partnership: the National Science Foundation (United States),
the Particle Physics and Astronomy Research Council (United Kingdom),
the National Research Council (Canada), CONICYT (Chile), the Australian
Research Council (Australia), CNPq (Brazil), and CONICET (Argentina).
The \nir\ image in Figure 2b is based on observations with the European
Southern Observatory telescopes obtained from the ESO/ST-ECF Science Archive
Facility.

\clearpage

\begin{figure}
\plotone{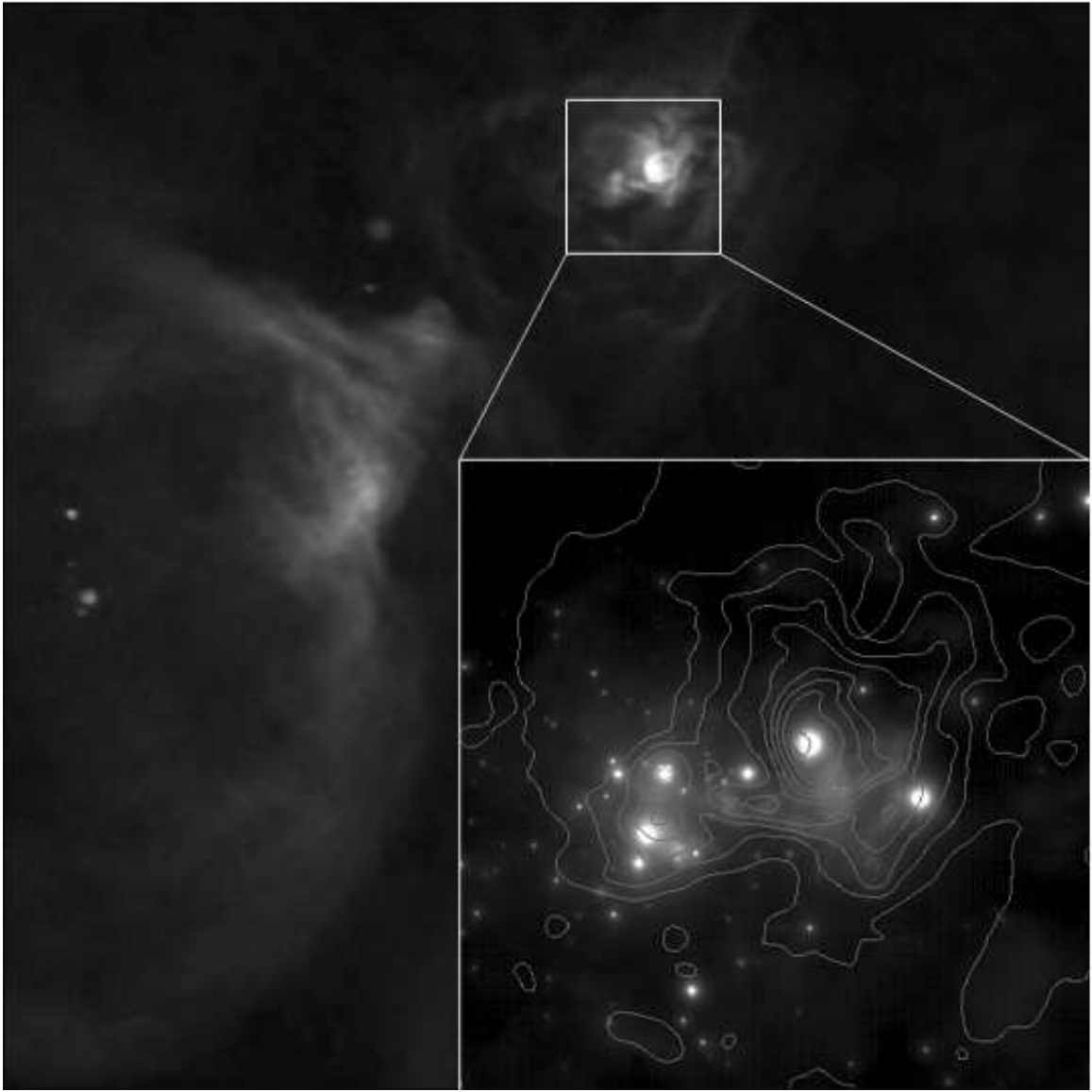}
\caption{a) (background image)
3.6\,cm free-free continuum emission from W51A (IRS\,1 and IRS\,2).
IRS\,2 is near the top of the image.
The image was made from the VLA ABCD configuration data set described in
\citet{mehringer94}.
The grayscale uses a square-root stretch.
The image extent is $112^{\prime\prime} \times 112^{\prime\prime}$.
b) (inset) H-band continuum emission from stars and ionized gas
in IRS\,2, with 3.6\,cm contours superimposed.
The H-band data were obtained from the ESO VLT archive.
The grayscale and contours use a square-root stretch.
The image extent is $15^{\prime\prime} \times 15^{\prime\prime}$.}
\end{figure}

\clearpage

\begin{figure}
\epsscale{0.8}
\plotone{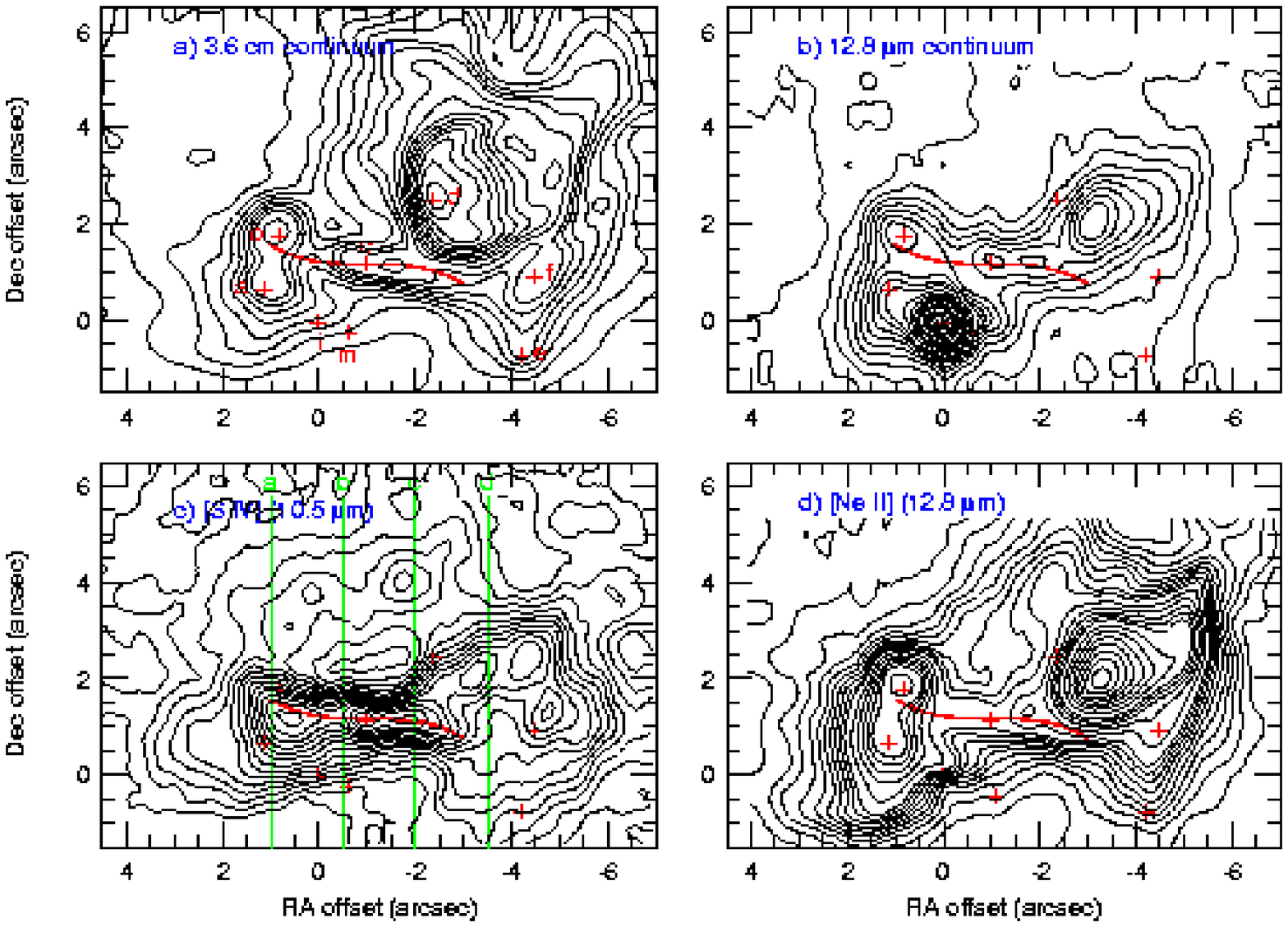}
\caption{Radio free-free, infrared continuum, and infrared line maps
of W51 IRS\,2.
Axes are labeled in arcseconds from the infrared continuum
source IRS\,2i at 19\hr 23\mt 40.10\st\ 14\degree 31\am 06.0\as\ (J2000).
a) 3.6 cm emission from Mehringer (1994), with various H\,II and maser
peaks marked.
Contours are at 1,2,...8,10,...26 $\times$ 0.0025 Jy\,beam$^{-1}$
or $7.5 \times 10^9$ Jy\,sr$^{-1}$.
H\,II peaks marked a-f are referred to in the text as IRS\,2a-f.
The curve through c shows the region of blue-shifted emission.
IRS\,2i is the infrared continuum source, and m is the maser center.
b)  12.8\,\um\ continuum emission.
Contours are at 1,2,...10,12,...20,24,...44
$\rm \times\,0.014\,erg / s\,cm^2\,sr\,cm^{-1}$.
c)  \siv\ (10.5\,\um ) line emission.
Contours are at 1,2,...15
$\rm \times\,0.009\,erg / s\,cm^2\,sr$.
Lines show locations of P-V cuts in Figure 3.
d)  \neii\ (12.8\,\um ) line emission.
Contours are at 1,2,...10,12,...20,24,...48
$\rm \times\,0.023\,erg / s\,cm^2\,sr$.}
\end{figure}

\clearpage

\begin{figure}
\epsscale{0.8}
\plotone{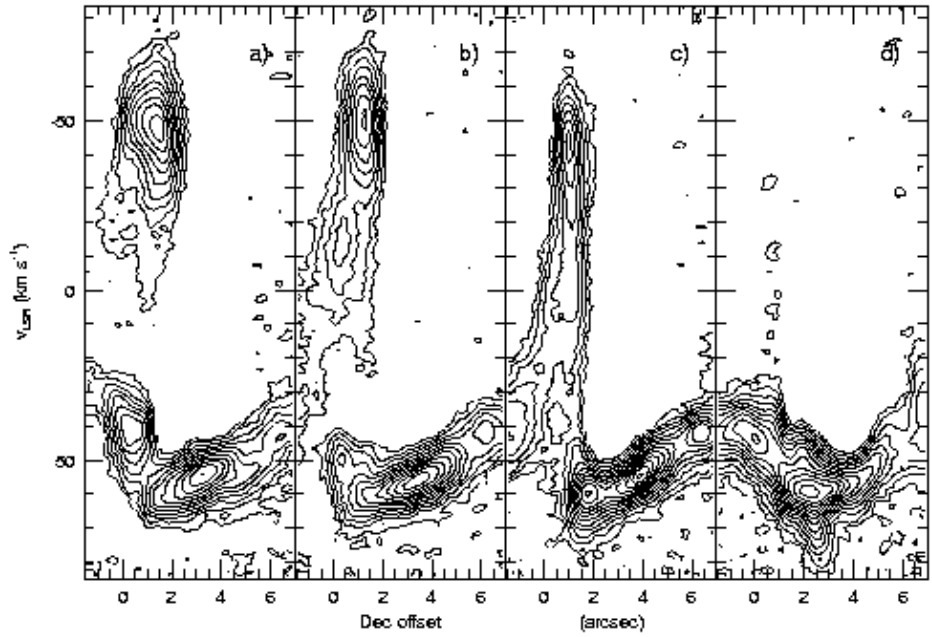}
\caption{Position-velocity cuts through the \siv\ data cube.
Horizontal axes are labeled in arcseconds north from IRS\,2i.
The contours are spaced quadratically.
The ambient molecular cloud is near \vlsr\ = 57\,\kms .
Locations of cut lines are shown in Figure 2c.
The online movie shows the data cube as a sequence of P-V diagrams.
The movie frames are labeled by the R.A. offset from IRS\,2i.}
\end{figure}

\clearpage

\begin{figure}
\epsscale{.35}
\plotone{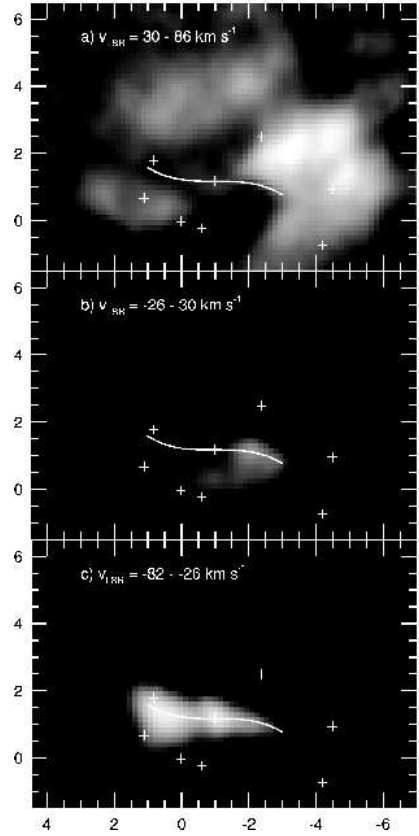}
\caption{Images of [S IV] emission in three velocity intervals.
The brightness stretch is linear and the same for all images.
Annotations are as in Figure 2.
a) Ionized gas near the molecular cloud velocity.
We interpret the dark region at bottom center as the region where
the cloud surface gas has been disrupted by the jet.
b) The `bridge' gas that we interpret as having been accelerated by the jet.
c) Blue-shifted emission from the jet which has been ionized
by stellar UV radiation.
The online movie show the data cube as a sequence of 180 narrow-band images,
running from \vlsr\ = +85\,\kms\ to -82\,\kms }
\end{figure}

\end{document}